\documentclass[aps,epsf]{revtex4}   

\usepackage{graphicx}
\newdimen\epsfxsize
\newdimen\epsfysize
\newcommand\epsfbox[1]{\includegraphics[width=\epsfxsize]{#1}}

\begin{document}

\title{Exploring proton rich systems and Coulomb induced instabilities}

\author{S.J. Lee}
\address{Department of Physics and Institute of Natural Sciences,
              Kyung Hee University, Suwon, KyungGiDo, Korea}

\author{A.Z. Mekjian}
\address{Department of Physics, Rutgers University, Piscataway NJ 08854}



\begin{abstract}

The thermodynamic properties of proton rich systems 
are explored in a mean field approach
which is generated from a Skyrme interaction.
The addition of Coulomb interactions result in asymmetries
which modify the chemical and mechanical instability of
the system and its equilibrium properties.
These properties are studied for systems with proton 
fraction $y$ on the proton richer side of the valley 
of $\beta$-stability as well as the neutron rich side.
Coulomb induced instabilities lead to proton diffusion processes 
on the proton richer side and also large asymmetries in
chemical and mechanical instabilities and coexistence curves.
Considering the whole range of $0 \le y \le 1$, 
we can study how the symmetry about $y=1/2$ is broken 
by asymmetric interaction and 
we can also explicitly show that the role between proton and neutron
is exchanged around $y_E$.
It is shown that there are 
two asymmetric coexistence surfaces in $(y, P, T)$ space, one for $y < y_E$ 
and another for $y > y_E$ and touching each other at $y_E$.
These asymmetries in instabilities show up as new branches,
one for $y < y_E$ and one for $y > y_E$, 
and thus form a closed loop in pressure versus $\rho$ for both 
chemical instability and coexistence regions. 
The branch of $y > 1/2 > y_E$ was not previously investigated
since only the $y < 1/2$ region is usually considered.
In our simplified model,  
mechanical instability is still symmetric around a 
point $y_E \ne 1/2$ even with Coulomb forces present.

\end{abstract}


\maketitle


Understanding properties of nuclei with large asymmetry in the 
proton neutron ratio has had a renewed interest for several reasons.
For example, future experiments done with rare isotope accelerators (RIA)
can explore properties of nuclei far from the valley of $\beta$-stability
on both the neutron and proton rich sides of it.
These studies would be useful for understanding the production of
the elements and neutron stars in the area of nuclear astrophysics.
Current interest stems from the liquid-gas phase transition \cite{adnp26}
which is experimentally studied using medium energy heavy ion reactions.
A primary goal of heavy ion reactions is to explore the phase
diagram of hadronic matter both at low and high density and
temperature looking for signals of phase transitions.
Another reason for studying asymmetric two component nuclear systems is 
that they offer a unique example of the 
thermodynamic and statistical properties of binary systems.
Here, the nuclear phase diagram is governed by nuclear volume,
surface, symmetry energy and asymmetric Coulomb effects.
The interplay of these on the phase structure has been of
interest in past studies \cite{prc63,prc68}.
Examples of other two component systems are binary alloys
and liquid $^3$He.
For $^3$He the two components are spin up and spin down fluids
and important new phase structure appears within the superfluid
phase itself below 3 mK.

In this paper we extended our previous study \cite{prc63,prc68} of phase 
structure of nuclear systems below the liquid-gas critical point to include
the entire range of the proton fraction $y = Z/(N+Z)$ from 0 to 1.
Our previous study focussed on $0 \le y \le 1/2$.
For systems with no asymmetric Coulomb effects,
a symmetry exists about the $y = 1/2$ point.
This symmetry is broken by the asymmetric Coulomb term.
Due to this symmetry, usually only $y < 1/2$ side was considered
and thus such studies would miss various symmetry breaking features and
the existence of another pair of coexistence points at large $y$.
Some preliminary initial new findings were presented in
Ref.\cite{prc63,prc68} in a restricted region of $y \le 1/2$
which are further elaborated on here and also extended 
to the whole range of $0 \le y \le 1$.
In particular we will focus on the proton rich side of 
the valley of $\beta$-stability and study the chemical and
mechanical instabilities and equilibrium properties of these systems.
Several previous mean field studies of two component systems
were carried out without the Coulomb force.
The pioneering work of M\"uller and Serot \cite{muller}
focussed on the importance of the nuclear symmetry energy
on the equilibrium structure of liquid-gas phase transition.
V. Baran etal \cite{baran} investigated instability properties
of such systems and pointed some new properties of the
liquid-gas phase transition associated with chemical instability
and isospin distillation based on symmetry energy considerations.
Calculations were extended by Colonna etal \cite{coloprl88}
to finite nuclei including surface and Coulomb effects.
Their calculation focussed on the instability modes and their
results showed that an octupolar mode dominates.
Pawlowski \cite{pawlow} has also recently considered the
importance of Coulomb and surface effects on the equilibrium
phase structure of the liquid-gas phase transition and found
them to be important.
An extensive summary of isospin effects in heavy-ion collisions 
at medium energy can be found in the book of Bao-An Li
and Schr\"oder \cite{baoan}.


In our study we will use a simplified Skyrme interaction to
generate the nuclear long range attraction, short range repulsion
and also the nuclear symmetry energy.
We include a simplified Coulomb energy which generates a
proton-neutron asymmetry.
The resulting interaction energy density is 
\begin{eqnarray}
 U(\rho) = - a_0 \rho^2 + a_3 \rho^3 + a_S (2y-1)^2 \rho^2 + C y^2 \rho^2
     \label{poten}
\end{eqnarray}
The $a_0$ term is the long range attraction which is quadratic
in the density $\rho$.
Shorter range repulsion is the $a_3$ term and we take
its density dependence to be $\rho^3$.
The nuclear symmetry energy is the $a_S(2y-1)^2$ term 
with $a_S = \frac{2}{3} (\frac{1}{2} + x_0) a_0$ and
has a $\rho^2$ dependence.
The asymmetric Coulomb term is $C y^2 \rho^2$.
The $C$, which in principle depends on nuclear size $R$, 
is taken to be constant.
For this choice, we can consider $R$ to be an effective
range of the Coulomb interaction.
Then, in this simplified limit, the Coulomb interaction depends
on the square of the proton density.
Our choice for the Coulomb energy and nuclear interaction
energy greately simplifies the full complexity of the analysis
and results in analytical solutions for instability boundaries.
By using a somewhat simplified imteraction,
we hope to qualitatively illustrate the main physical phenomena
associated with phase transitions in two component systems
which include both symmetry and asymmetric terms.
We also believe that the main results will not change qualitatively,
nor in any major quantitative way when more 
complicated dependences are included.

\begin{figure}

\centerline{ \epsfxsize=5in \epsfbox{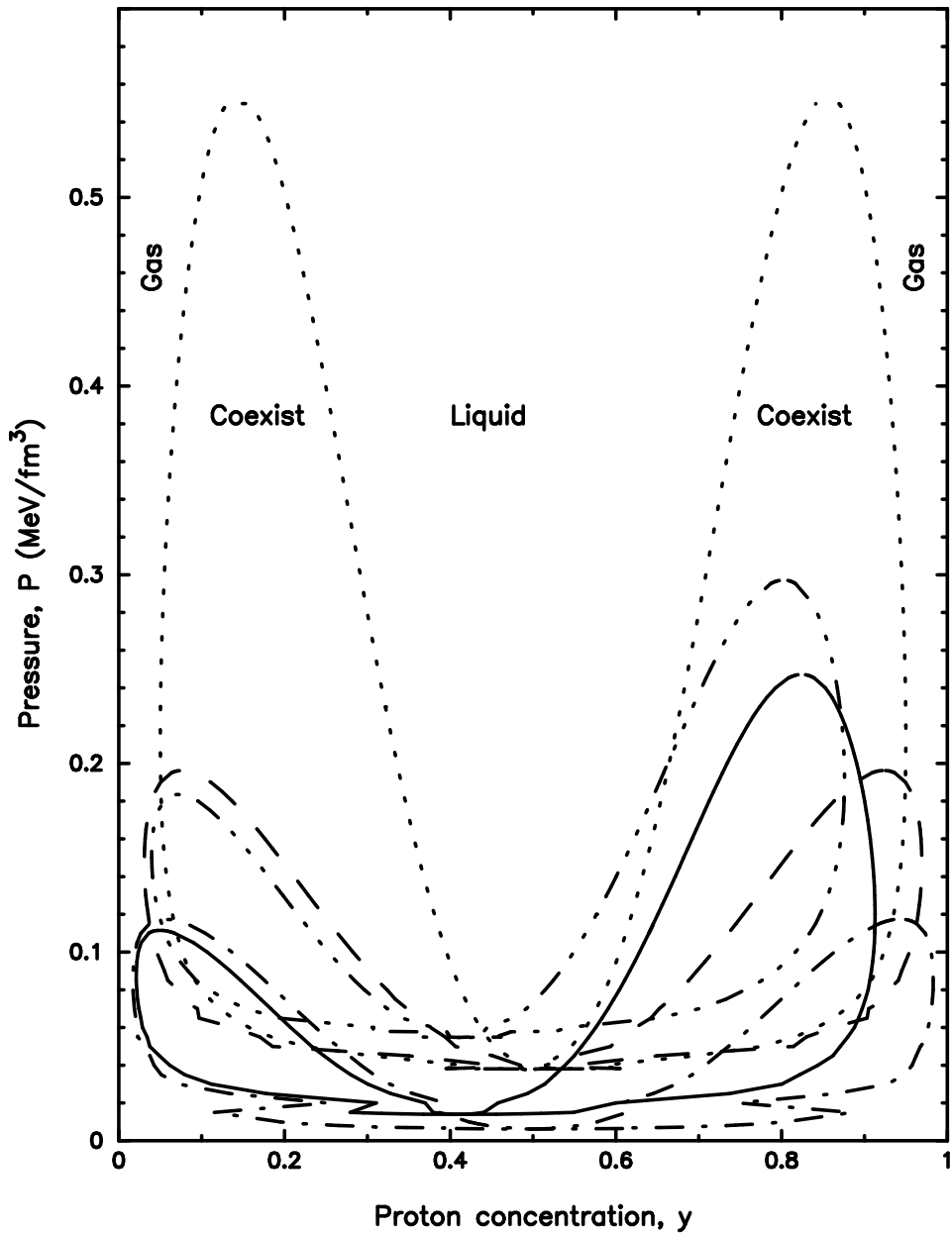} }

\caption{
Pressure $P$ versus proton fraction $y$ for coexistence loop at $T = 10$ MeV.
The solid line is for the case with Coulomb and surface effects.
The dash-dotted line has surface, but no Coulomb terms.
Coulomb effects are shown in the dash-dot-dot-dotted line
with no surface terms.
The dashed line is for the case without Coulomb and surface terms
while the dotted line is for a larger symmetry energy.
The regions of liquid and gas phases and the coexistence are
indicated explicitely for the dotted curve.
  } \label{fig1}
\end{figure}

\begin{figure}

\centerline{ \epsfxsize=5in \epsfbox{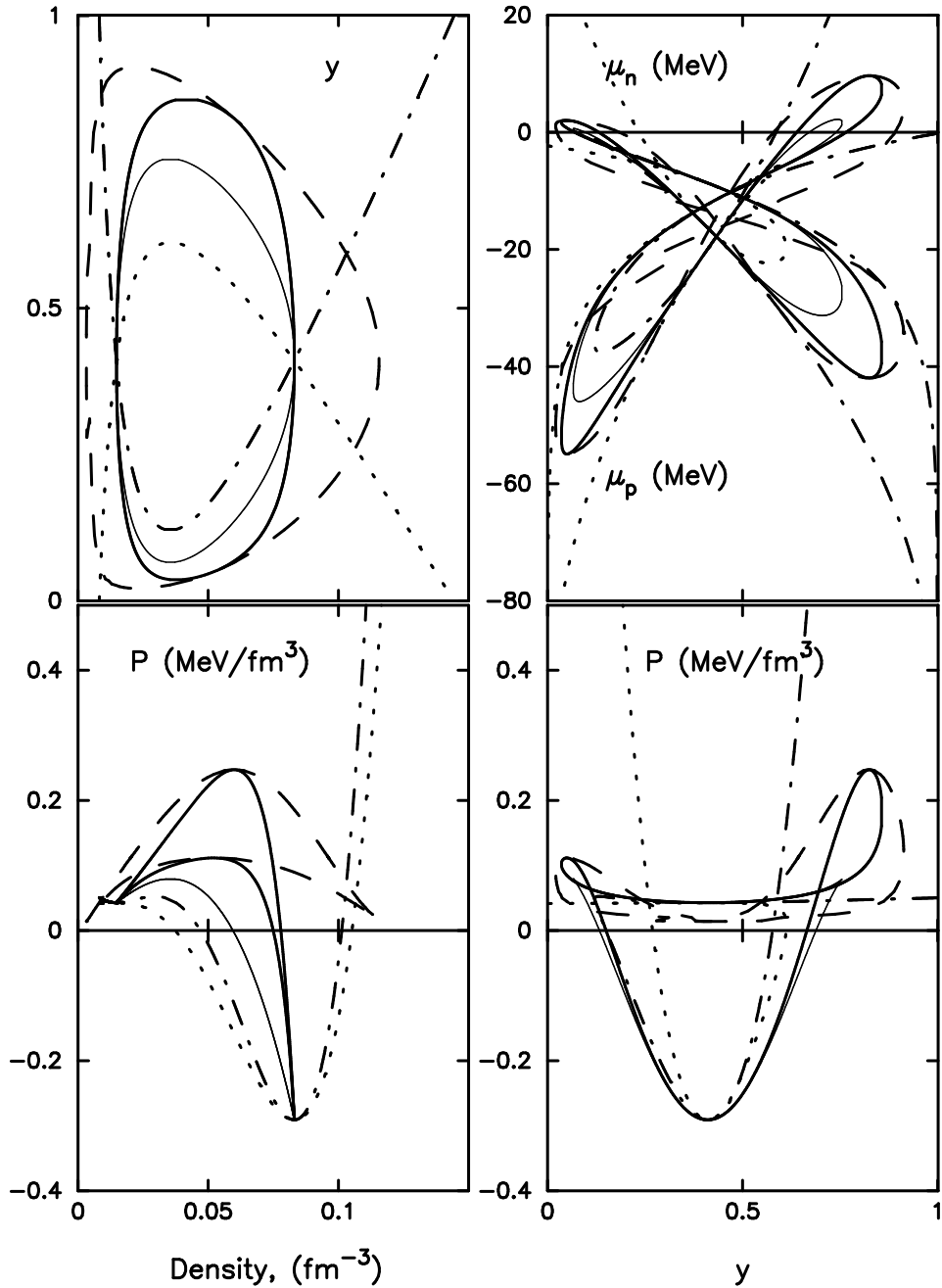} }

\caption{
Figures show the coexistence curves (dashed line),
chemical instability curves (thick solid line) and
mechanical instability curves (thin solid line).
Also shown are the $\partial\mu_q/\partial\rho = 0$ curves
for proton (dotted line) and for neutron (dash-dotted line)
at $T = 10$ MeV.
The Coulomb and surface terms are included here.
  } \label{fig2}
\end{figure}

A kinetic energy density is added to $U(\rho)$.
Since we will present results for moderate density and
temperatures of the order of 10 MeV, 
we will work to first order in the degeneracy corrections.
In this limit the pressure is
\begin{eqnarray}
 P = \rho T
    - \left(a_0 - \frac{T}{2\sqrt{2}} \frac{\lambda^3}{4\gamma}\right) \rho^2
    + 2 a_3 \rho^3
    + \left(\frac{2}{3} (\frac{1}{2} + x_0) a_0 + \frac{T}{2\sqrt{2}}
          \frac{\lambda^3}{4\gamma}\right) (2y-1)^2 \rho^2
    + C y^2 \rho^2        \label{pres}
\end{eqnarray}
The $a_0 = - \frac{3}{8} t_0$ and $a_3 = \frac{1}{16} t_3$
with $t_0 = -1089$ MeV fm$^3$ and $t_3 = 17480.4$ MeV fm$^6$
as shown in Ref. \cite{prc63}.
The $\gamma = 2$ is the spin degeneracy factor and
$\lambda = h/(2\pi m T)^{1/2}$.
The proton and neutron chemical potentials are
\begin{eqnarray}
     \mu_q &=& T\ln[(\lambda^3/\gamma)(1 \pm (2y-1))\rho/2]
           - 2a_0 \rho + 3a_3\rho^2 
                  \nonumber  \\  & &
     \pm (4/3)(1/2 + x_0) a_0 (2y-1)\rho 
   + (1 \pm 1) C y \rho
   + (T/2\sqrt{2}) (\lambda^3/\gamma) [1\pm(2y-1)]\rho/2
          \label{chem}
\end{eqnarray}
with the upper sign for protons and the lower sign for neutrons.
These three equations determine the equilibrium phase structure and
the mechanical and chemical instability regions of our simplified
two component nuclear system.
The phase structure is a three dimensional surface in $y$, $T$ and $P$.
At $y = 1/2$, symmetric systems with no Coulomb interaction, 
one has the familiar $P$, $T$ curve of Maxwell pressure versus $T$ 
for equilibrium state which ends at the critical point.
For fixed $T$, the equilibrium surface intersects the fixed $T$ 
plane to form loops in $P$ versus $y$.
These loops represent the proton fraction in the liquid and
gas phases which are different except at a point called $y_E$
which is not 1/2 for the case with Coulomb interaction.
These equilibrium loops end at the critical points $(y_c(T), P_c(T))$
at fixed $T$ or $(P_c(y), T_c(y))$ at fixed $y$
having the same densities both in liquid and gas phases.
The point with lowest $y$ value on this loop is the maximally asymmetric point
at a fixed $T$.
Another maximal asymmetric point is at $y > y_E$ with largest $y$ value.
The equilibrium surface is generated from the chemical
potentials $\mu_q(y,T,P)$ by requiring at fixed $P$ and $T$
the neutron chemical potential in liquid and gas phase to be
equal and proton chemical potential in these two phases to be equal
at the same two values of $y$.
Geometrically, this corresponds to the construction of a 
rectangular box in a plot of $\mu$ versus $y$
at fixed $P$ and $T$ where the horizontal lines are the 
same neutron chemical potentials and the same proton chemical potentials 
and the vertical lines are the two values of $y$ where each chemical
potential in liquid and gas phases are equal.
The gas phase will be more asymmetric and have lower $y$ for $y < y_E$ 
and larger $y$ for $y > y_E$ than the liquid phase.
The liquid phase is more symmetric because of the symmetry energy term
and has the value of $y$ closer to $y_E$.
Figures \ref{fig1} and \ref{fig2}
shows various properties associated with the equilibrium surfaces
and behavior of the chemical potentials.

The point where the proton fraction is equal in both gas and liquid
phase can be obtained from the condition $\partial P/\partial y = 0$ and
this point is given by
\begin{eqnarray} 
 y_E = \left(\frac{1}{2}\right)
    \left/\left(1 + \frac{C/4}{[(\frac{2}{3}) (\frac{1}{2}+x_0) a_0
     + \frac{T}{2\sqrt{2}} \frac{\lambda^3}{4\gamma}]}\right)\right. \label{ye}
\end{eqnarray}
An important feature of the Coulomb force is to shift $y_E$ away
from its value of 1/2 when $C = 0$ to a value closer to the valley
of $\beta$-stability found from $d (E/A)/d\rho = P = 0$ at zero $T$.
This shift is important for understanding the phenomena of isospin
fractionization which favors more symmetric liquids and less
symmetric gases.

Fig.\ref{fig1} shows the effect of various terms on the coexistence 
curve obtained from the intersection of the coexistence surface
with a fixed $T = 10$ MeV plane. 
Here we have used the same parameters as used in Ref.\cite{prc63,prc68}
except for using a more realistic smaller symmetry energy ($x_0 = -1/6$).
The curve has two symmetric loops which are connected with 
minimum $P$ at $y = 1/2$ when Coulomb forces
are turned off and the loops are larger for a large symmetry
energy (dotted line with $x_0 = 1/2$) 
compared to smaller symmetry energy (dashed line with $x_0 = -1/6$).
A comparison of these two curves show that the symmetry energy has
essentially no effect on the minimum point at $y = 1/2$ 
(symmetry energy is zero at this $y$)
while the maximum points (which are not at $y = 1/2$) 
has larger pressure for a larger symmetry energy.
The dash-dotted line with minimum at $y = 1/2$ includes a surface
energy or surface tension term besides a symmetry energy term
whose value is the same as that for the dashed line.
By comparing these two loops, the effect of a surface term
is to shift the coexistence surface to lower pressure.
The loops are still symmetric around $y = 1/2$ and
slightly compressed by the droplet's surface tension.
The shifting by the surface term can also be seen from other curves
with Coulomb terms included.
The solid curve contains the symmetry energy, surface energy,
and Coulomb energy while the dash-dot-dot-dotted curve has
symmetry and Coulomb effects.
The main effects of the Coulomb term are:
1) to shift the intersection point of the two loops to $y_E = 0.41$  
for the parameters considered and to higher pressure and
2) to make the loops very asymmetric with the coexistence loop
for $y < y_E$ becoming smaller and 
the coexistence loop for $y > y_E$ becoming larger.
Most of the loop for $y > y_E$ was not studied and thus was not closed
in previous studies \cite{prc63,prc68} due to the restriction of $y \le 1/2$.
The existence of coexistence pair with more proton in gas
than in liquid was first pointed out in Ref.\cite{prc63}
but in a small window of pressure since we restriced $y \le 1/2$
which allow only up to $y=1/2$ for the gas phase.
For each case, two loops intersect at $y_E$ 
which occurs at the lowest pressure
while the maximum pressure for each loop has a different $y$ value.
The liquid and gas coexist inside of these two loops
since the boundaries are the pure liquid and gas phases.
The liquid and gas phases are separated outside of these loops;
liquid phase for $y$ value between two maxima and
gas phase for $y$ value outside the two maxima.
These are indicated explicitly for the dotted curve.
A system with $y < y_E$ inside the coexistence region is
composed of a proton richer liquid and a proton deficit gas
compared to the original $y$
while a system with $y > y_E$ in the coexistence region
has more protons in gas phase than in liquid component.
The Coulomb term shrinks the coexistence region for $y < y_E$
while it enlarges the coexistence region for $y > y_E$ (Fig.\ref{fig1}).

The boundaries for mechanical and chemical instability are
easy to generate.
Of these two instabilities, the region of mechanical instability is
somewhat easier to obtain and follows from $(dP/d\rho)_{y,T} = 0$
which leads to a quadratic equation, $a \rho^2 + 2 b \rho + c = 0$.
Using Eq.(\ref{pres}), this condition gives
\begin{eqnarray}
 \rho_{\pm} &=& \frac{- b \pm \sqrt{b^2 - a c}}{a}  \\
 a &=& 6 a_3       \nonumber  \\
 b &=& - \left(a_0 - \frac{d}{4}\right) 
        + \left(\frac{2}{3} (\frac{1}{2} + x_0) a_0 
             + \frac{d}{4}   
           \right) (2y-1)^2 
        + C y^2     \nonumber  \\
 c &=& T
        \nonumber
\end{eqnarray}
where $d = \frac{T}{2\sqrt{2}} \frac{\lambda^3}{\gamma}$.
The presence of $d$ represents a first order quntal correction to
the non-degenerate limit. The Coulomb term and symmetry energy
terms appear in $b$.
The critical point $(y_c(T), P_c(T))$ or $(P_c(y), T_c(y))$ occurs 
when $b^2 = a c$ with the critical density $\rho_c = -b/a$.

We note that the coefficient $b$ can be rewritten in a form
involving $b = d_1 (y-y_E)^2 + d_2$ with
 $d_1 = \left(\frac{8}{3} (\frac{1}{2}+x_0) a_0 + d \right) + C$ and
 $d_2 = \frac{C}{4}/
      \left(1 + C/[(\frac{8}{3}) (\frac{1}{2}+x_0) a_0 + d]\right)
     - (a_0 - \frac{d}{4})$.
Thus $b$ is symmetric around $y_E$ and since $y$ does not appear
in $a$ or $c$, the solution $\rho_\pm$ as a function of $y$
is also symmetric around $y_E$.
Therefore, the mechanical instability curve of $P$ versus $\rho$ will
have a symmetry that results in one curve instead of a loop.
Also the mechanical instability loop in $P$ versus $y$ 
and $y$ versus $\rho$ are symmetric around $y_E$.
This fact is not true for chemical instability and the coexistence curve.
These facts will be shown in Fig. \ref{fig2}.

The boundary of chemical instability can be obtained 
from $(d\mu_q/dy)_{P,T} = 0$.
These conditions for proton and neutron give the same relation
since $y d\mu_p + (1-y) d\mu_n = (1/\rho)dP$.
For two component system with a fixed total fraction, if one component changes
from one phase to other then the other component must compensate this concentration
change since they are not independent of each other.
The chemical instability condition can be rewritten in terms of derivatives
of chemical potential and pressure with respect to the
variables $\rho$ and $y$ which appear in Eq.(\ref{pres}) 
and Eq.(\ref{chem}).
Namely, the chemical instability boundary can be obtained from
$(dP/d\rho)_{y,T} = (dP/dy)_{\rho,T} (d\mu_q/d\rho)_{y,T}/(d\mu_q/dy)_{\rho,T}$.
Then using the equation of state and chemical potential equation,
the chemical instability boundary is determined by solutions
to a cubic equation
\begin{eqnarray}
  & & \hspace{-1cm}
 c_3\rho^3+c_2\rho^2+c_1\rho+c_0=0    \\
 c_3 &=& \left[32 (\frac{1}{2}+x_0) a_0 + 12 C + 12 d\right] a_3 y(1-y)
             \nonumber \\
 c_2 &=& 6a_3 T - \left[(\frac{32}{3}) (\frac{1}{2}+x_0) a_0^2
    + (\frac{8}{3}) (1-x_0) a_0 C
    - d^2 + (\frac{32}{3}) (1-x_0) a_0 d - 2 d C \right] y(1-y)   \nonumber \\
 c_1 &=& \left[-(\frac{4}{3}) (1-x_0) a_0 + 2Cy + d\right] T     \nonumber \\
 c_0 &=& T^2  \nonumber
\end{eqnarray}
We note that the term asymmetric in $y \leftrightarrow (1-y)$ is
the coefficient $c_1$, while $c_2$ and $c_3$ are invariant 
under this interchange.
The solution to the cubic has three roots.
For a physical situation, two roots will be positive and physical
and the third negative and unphysical.
Since $c_0$ and $c_3$ are positive, 
the product of the 3 roots is negative.
For each $y$, the solution to the cubic generates a chemical 
instability curve in $\rho$ and $T$
and a three dimensional surface in $\rho$, $y$, $T$.
Each point on this surface has corresponding values of $P$ and $\mu_q$.
Cutting the $(\rho,y,T)$ surface of chemical instability with a fixed $T$ 
plane gives a $\rho$, $y$ curve for the chemical instability
and on this curve $P$ and $\mu_q$ are specified.
Then using this fact, we can generate $\mu_q$ versus $y$,
$P$ versus $\rho$, and $P$ versus $y$ curves for the chemical
instability shown in Fig.\ref{fig2}.
Similar mechanical instability and coexistence surfaces can
be generated. These results are also shown in Fig.\ref{fig2}.

Figure \ref{fig2} contains plots of $y$ versus $\rho$,
the loops of $\mu_p$ and $\mu_n$ versus $y$,
curves of $P$ versus $\rho$ and curves of $P$ versus $y$.
Interesting points occur at $d\rho/dy = 0$ on these curves
which correspond to the cusp points at $y_E$ in $P$ versus $\rho$
which have two different branches for the coexistence curve and
chemical instability curve. 
The mechanical instability region has only one curve,
which starts and ends at $y_E$,
because of the symmetry of it around $y_E$.
The upper curve corresponds to the proton richer side with
respect to $y_E$ and the lower branch is for the neutron
richer side of $y_E$.
These two curves form a closed loop in a constant $T$ plane.
The upper branch did not appear in Ref.\cite{prc68}
due to the restriction of $y \le 1/2$.
When the Coulomb force is turned off 
these two branches in the coexistence curve and
chemical instability curve would collapse into one curve also
due to the symmetry around $y = 1/2$.
The maximal asymmetry points correspond to $dy/d\rho = 0$,
one at low $y$ and one at high $y$.
In the $y$ versus $\rho$ curve, the mechanical instability loop
is inside the chemical instability loop except at two points 
where it is tangent to it.
The tangent is at $y_E$.
The chemical instability loop in $y$ versus $\rho$ is inside
the coexistence loop and tangent to it at two points, 
one at $y < y_E$ and one at $y > y_E$.
These values of $y$ correspond to the critical $y_c(T)$,
points with maximum pressure of each branch 
having the same densities both in liquid and gas phase,
at a fixed temperature, 10 MeV for Fig.\ref{fig2}.
For $P$ versus $y$ the mechanical instability loop is
symmetric about $y_E$, while the chemical and coexistence curves are
not when Coulomb forces are present,
but are symmetric when Coulomb forces are turned off.
The Coulomb instability leads to proton diffusion above $y_E$
and neutron diffusion below $y_E$.
The chemical potential $\mu_q$ and the $\partial\mu_q/\partial\rho = 0$
curves show that the role between proton and neutron is exchanged
around $y_E$.
In Ref.\cite{prc68}, the $\partial\mu_p/\partial\rho = 0$ curve
had two disconnected portions due to the cut at $y = 1/2$ and 
was quite different from the $\partial\mu_n/\partial\rho = 0$ curve.

Similar calculations as shown in Fig.\ref{fig2} for the
cases without Coulomb and/or surface terms can also be done 
even though we did not included the figure in this paper to
keep the fiqure simple. 
The effects of each term when treated separately are as follows. 
The surface energy brings the pressure down, 
increase the chemical potential of the upper part of each loop, 
and slightly enlarges the loops in $\rho$-$y$ plot.
The Coulomb term breaks the $y \leftrightarrow (1-y)$ symmetry
and shrink the $\rho$-$y$ plot to the lower density and proton
concentration while increasing the lower part of the chemical potential
and pressure curves.

In summary, in this paper we investigated properties
of the coexistence region, and mechanical and chemical instability
regions when asymmetric Coulomb forces are also included besides
volume, surface, and symmetry energy terms in a mean field
description of hot nuclear matter for the whole range of proton 
fraction $0 \le y \le 1$ 
(only the range of $0 \le y \le 1/2$ was considered previously
for the coexistence region \cite{prc63,prc68} and
for the instability region \cite{prc68}).
Including the Coulomb interactions leads to a changes in many features 
that are physically important.
These include a shift in $y_E$, the value of the proton fraction at
which the liquid and gas phases would coexist at the same proton fraction 
from $y = 1/2$ to a value of $y$ closer to the valley of $\beta$-stability.
The shifting of $y_E$ from 1/2 to a lower value by the addition of
the Coulomb force is an important effect for isospin fractionization
or distillation \cite{ref9}.
An investigation of the chemical instability properties shows that
proton diffusion will occur above the value of $y_E$
while neutron diffusion occurs below it.
By extending our calculation to values of $y > 1/2 > y_E$
we were also able to show some new features of the coexistence
and instability regions. 
First, the constant $T$ cuts of the corresponding surfaces
now form closed loops while the constant $T$ cuts
of $\partial\mu_q/\partial\rho = 0$ surfaces form continuous curves.
Previously, the proton curve $\partial\mu_p/\partial\rho = 0$ had 
two disconnected pieces due to the restiction $y \le 1/2$ \cite{prc68}.
The main behavior of the temperature dependence of these surfaces 
remain the same as in Ref.\cite{prc68} with the role of proton and 
neutron exchanged in going from the $y < y_E$ region to the $y > y_E$ region.
The mechanical instability region in $P$ versus $\rho$
or $P$ versus $y$ is symmetric around the shifted $y_E$
when Coulomb interactions are included in our model.
This symmetry does not occur for the coexistence curve and 
chemical equilibrium curves.
Rather, these have new branches which appear in $P$ versus $\rho$
and the two branches cusp at the two $\rho$'s which have $y = y_E$.
The upper branch corresponds to the proton rich side of $y_E$, or $y > y_E$,
while the lower branch corresponds to the neutron rich side
of $y_E$, or $y < y_E$.
The coexistence curve now has two asymmetric loops in $P$ versus $y$
that meet at the shifted $y_E$. 
The upper branch of the instability curve with $y > 1/2$ and coexistence
loops at $y > y_E$ were missing in previous studies.
For systems with $y > y_E$, the gas phase can become proton richer from 
proton diffusion induced by a Coulomb instability of the liquid phase.
For $y < y_E$, the gas phase will be neutron richer from
neutron diffusion.
This paper also shows how the Coulomb interaction breaks some of
the symmetries associated with isospin in a hot nuclear system
and its associated liquid-gas phase transition.
The effects of a surface energy terms are discussed.
The surface tension associated with the presence of a liquid drop
shifts the coexistence curve to lower pressure.

This work was supported 
in part by the DOE Grant No. DE-FG02-96ER-40987
and in part by Grant No. R05-2001-000-00097-0 from the Basic Research Program 
of the Korea Science and Engineering Foundation.

\end{document}